\documentclass[conference]{IEEEtran}
\IEEEoverridecommandlockouts
\usepackage{cite}
\usepackage{amsmath,amssymb,amsfonts}
\usepackage{algorithmic}
\usepackage{graphicx}
\usepackage{textcomp}
\usepackage{xcolor}
\usepackage{graphicx}
\usepackage{subcaption}
\usepackage{makecell}
\usepackage{tabularx} % For better table handling
\usepackage{adjustbox} % For scaling tables if needed

\usepackage{amsmath}         % For math environments like \begin{equation}
\usepackage{amssymb}         % For additional math symbols
\usepackage{algorithm}       % For the algorithm floating environment
\usepackage{algorithmic}     % For algorithm pseudo-code formatting
\usepackage{enumitem}   

\usepackage{graphicx}
\usepackage{array}
\usepackage{booktabs}

\usepackage[a4paper, total={184mm,239mm}]{geometry}
\def\BibTeX{{\rm B\kern-.05em{\sc i\kern-.025em b}\kern-.08em
    T\kern-.1667em\lower.7ex\hbox{E}\kern-.125emX}}

\begin{document}
\raggedbottom

\title{LLM-USO: Large Language Model-based Universal Sizing Optimizer \\

}

\author{
\IEEEauthorblockN{1\textsuperscript{st} Karthik Somayaji N.S}
\IEEEauthorblockA{\textit{Dept. of Electrical and Computer Engineering} \\
\textit{University of California, Santa Barbara}\\
Santa Barbara, CA, USA\\
karthi@ucsb.edu}
\and
\IEEEauthorblockN{2\textsuperscript{nd} Peng Li}
\IEEEauthorblockA{\textit{Dept. of Electrical and Computer Engineering} \\
\textit{University of California, Santa Barbara}\\
Santa Barbara, CA, USA\\
lip@ucsb.edu}

}

\maketitle

\begin{abstract}

The design of analog circuits is a cornerstone of integrated circuit (IC) development, requiring the optimization of complex, interconnected sub-structures such as amplifiers, comparators, and buffers. Traditionally, this process relies heavily on expert human knowledge to refine design objectives by carefully tuning sub-components while accounting for their interdependencies. Existing methods, such as Bayesian Optimization (BO), offer a mathematically driven approach for efficiently navigating large design spaces. However, these methods fall short in two critical areas compared to human expertise: (i) they lack the semantic understanding of the sizing solution space and its direct correlation with design objectives before optimization, and (ii) they fail to reuse knowledge gained from optimizing similar sub-structures across different circuits. To overcome these limitations, we propose the Large Language Model-based Universal Sizing Optimizer (LLM-USO), which introduces a novel method for knowledge representation to encode circuit design knowledge in a structured text format. This representation enables the systematic reuse of optimization insights for circuits with similar sub-structures. LLM-USO employs a hybrid framework that integrates BO with large language models (LLMs) and a learning summary module. This approach serves to: (i) infuse domain-specific knowledge into the BO process and (ii) facilitate knowledge transfer across circuits, mirroring the cognitive strategies of expert designers. Specifically, LLM-USO constructs a knowledge summary mechanism to distill and apply design insights from one circuit to related ones. It also incorporates a knowledge summary critiquing mechanism to ensure the accuracy and quality of the summaries and employs BO-guided suggestion filtering to identify optimal design points efficiently. We evaluate the LLM-USO framework through transfer learning experiments on various analog circuits, demonstrating its ability to improve the quality of circuit design optimization.

\end{abstract}


\section{Introduction}

Analog circuit design plays a fundamental role in integrated circuit (IC) development, yet it remains one of the most complex and time-consuming tasks within electronic design automation (EDA) \cite{analog_time_consuming}. The intricacies of balancing multiple design objectives—such as power consumption, area minimization, and performance—combined with the variability in circuit topology and technology, necessitate a high degree of expertise from human designers. These experts typically rely on domain-specific knowledge and iterative optimization to achieve optimal designs.

In recent years, machine learning techniques, particularly Bayesian Optimization (BO) \cite{BO}, have emerged as promising tools for automating analog circuit design. BO’s ability to explore large, complex design spaces while efficiently trading off exploration and exploitation through probabilistic models has made it a popular choice in the field. However, despite its automation potential, traditional BO methods are not without limitations. As a black-box optimization technique, BO lacks an inherent understanding of domain-specific knowledge that could guide the search process more intelligently. Additionally, BO’s focus on optimizing individual circuits fails to leverage transferable knowledge from past designs, resulting in inefficient exploration and a lack of context reuse, particularly when optimizing circuits with similar sub-structures.

Large language models (LLMs) have recently been introduced to address some of these limitations by augmenting BO with their ability to understand and incorporate domain knowledge. The integration of LLMs enables more intelligent suggestions and a higher degree of contextual understanding during the design process. For example, the ADO-LLM framework \cite{ado-llm} combines BO and LLMs to generate high-quality initial design points and guide BO’s exploration using domain-specific insights. However, this framework still suffers from two major deficiencies. First, it does not utilize transfer learning, which in our context refers to reusing optimization knowledge from circuits with similar sub-structures. This omission prevents the framework from providing structured knowledge summaries from past optimizations as context for new tasks, thereby limiting its ability to accelerate optimization across related circuits. Second, while LLMs can suggest viable design points, their suggestions are often repetitive, leading to stagnation in the exploration of novel, promising regions and underutilization of the LLM’s design knowledge.

\begin{figure*}[!htbp]
  \centering
  \includegraphics[width=0.85\textwidth]{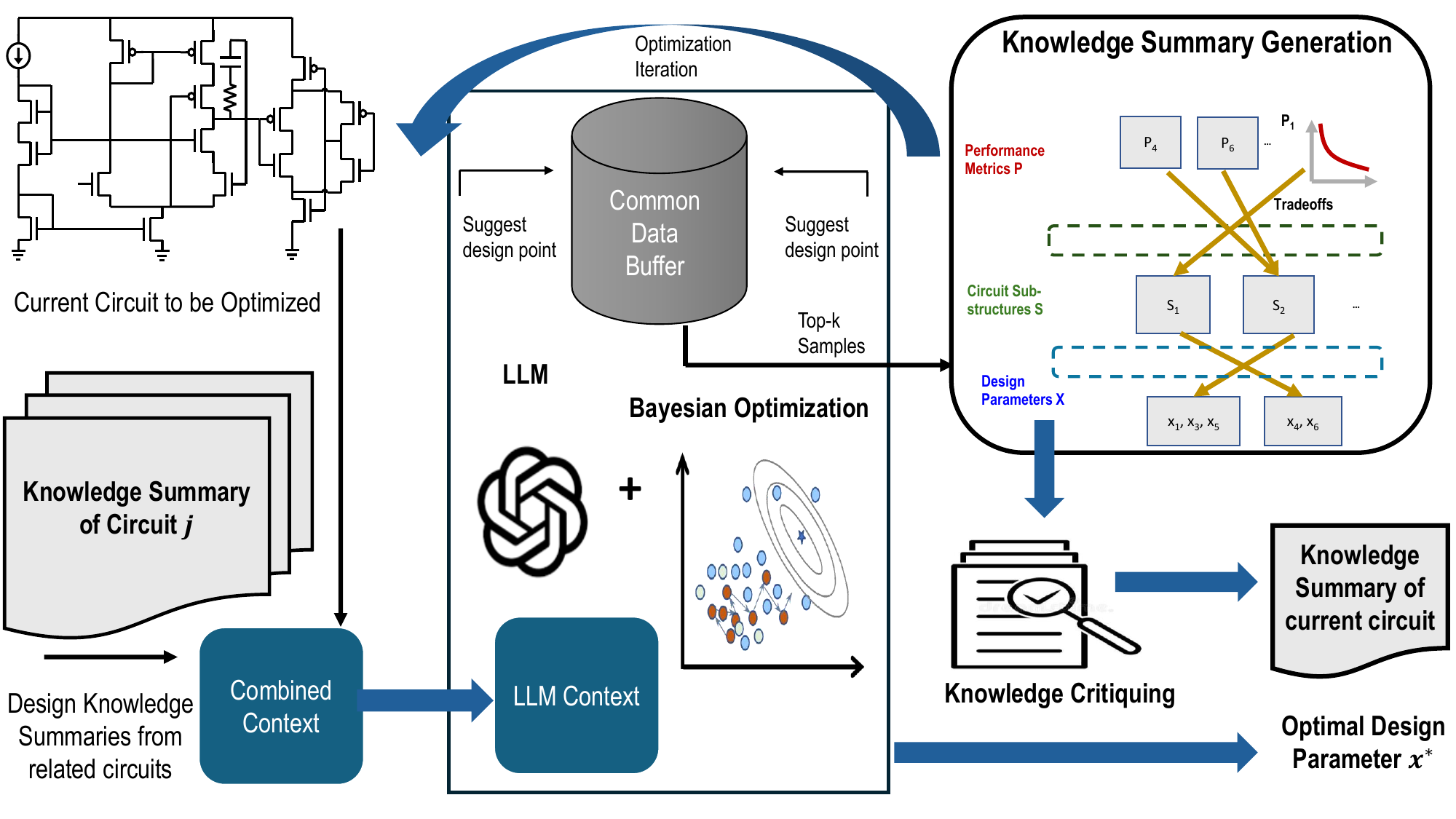}
  \caption{LLM-USO: Employs structured design knowledge summaries from related circuits for the optimization of the current circuit. LLM-USO generated the optimal design point $x^*$ and also the knowledge summary for the given circuit.}
  \label{fig:LLM-USO_flow}
 % \vspace{-2em}
\end{figure*}

To address these limitations, we first propose a novel method for representing analog design knowledge for a given circuit in a structured text format. This knowledge representation captures the interplay between circuit performance metrics, circuit sub-structures, and corresponding design parameters, providing a comprehensive framework for encoding and transferring design insights. This structured representation serves as the foundation for transferring optimization knowledge from one circuit to another, enabling efficient reuse of insights across circuits with similar sub-structures. Building on this, we propose a novel hybrid framework called the Large Language Model-based Universal Sizing Optimizer (LLM-USO). LLM-USO combines the strengths of BO and LLMs while overcoming the identified limitations. Central to LLM-USO is an information reuse mechanism that enables the transfer of optimization knowledge between circuits with shared sub-structures. By systematically extracting and reusing structured design knowledge from previously optimized circuits, LLM-USO mirrors the strategies employed by expert human designers, who rely on prior experience to optimize new designs more efficiently.

Additionally, LLM-USO leverages LLMs’ domain knowledge and BO's uncertainty estimates to intelligently explore the design space by ranking and filtering LLM-generated suggestions based on design point novelty. This synergy ensures that LLM-USO not only identifies promising new design points but also avoids redundant explorations, further enhancing efficiency. By integrating information reuse among related circuits with guided exploration, this hybrid approach improves the quality and efficiency of the optimization process. The LLM-USO framework is illustrated in Figure \ref{fig:LLM-USO_flow}. Thus our contributions include:

\begin{enumerate}
     \item \textbf{Knowledge Representation for Analog Circuits:} We propose a novel method to represent analog circuit design knowledge in a structured format for a given circuit. This representation captures the interplay between performance metrics, circuit sub-structures, and design (sizing) parameters, forming the foundation for effective knowledge transfer across circuits with similar sub-structures.

    \item \textbf{Structured Knowledge Summary Generation:} We propose a method to elicit the analog design knowledge representation during optimization iterations using a structured prompt design. This mechanism, integrated into the LLM-USO framework, efficiently summarizes design knowledge gained from optimizing a circuit. These summaries enable the reuse of insights in optimizing other circuits with similar sub-structures, mirroring expert human design strategies and enhancing efficiency.

 \item \textbf{Knowledge Critiquing Mechanism:} To ensure high-quality and complete design insights, we introduce a critiquing mechanism that rigorously evaluates and refines the generated knowledge summaries.
 
\item \textbf{Ranking LLM Suggestions with BO Uncertainty:} We develop an approach that integrates Bayesian Optimization uncertainty estimates to rank LLM-generated suggestions. This ranking prioritizes unexplored, high-potential design points, improving the efficiency and effectiveness of the optimization process.

\item \textbf{Comprehensive Evaluation and Performance Gains:} We validate the LLM-USO framework on multiple analog circuits, demonstrating significant improvements in design quality and efficiency through information reuse and intelligent ranking of LLM suggestions based on BO uncertainty.

\end{enumerate}

\section{Related Work}

Analog circuit design optimization has increasingly benefited from machine learning approaches, particularly reinforcement learning (RL) and Bayesian optimization (BO), given the complexity and high dimensionality of the design space. In RL, methods such as \cite{prl} employ prioritized sampling of replay buffers for more effective exploration, while \cite{autockt} utilizes algorithms like proximal policy optimization \cite{ppo} to optimize stochastic policies efficiently. Additionally, \cite{gcn_rl} integrates graph neural networks to enable transfer learning for circuit optimization across different technology nodes.

Bayesian optimization \cite{BO}, on the other hand, is widely adopted for its ability to balance exploration and exploitation in expensive simulations. Techniques like \cite{batch_BO} employ multiple acquisition functions to suggest design points, taking into account trade-offs between different acquisition strategies. Meanwhile, \cite{parallel_BO} improves the efficiency of BO through batch querying and parallel execution.

Recently, large language models (LLMs) have shown promise in analog circuit design automation. LADAC \cite{LADAC} automates design parameter selection using LLM-driven suggestions. In-context learning \cite{in_context_learning} and chain of thought prompting \cite{chain_of_thought} have enhanced LLMs’ ability to reason through complex design trade-offs and generate high-quality solutions. More recently, ADO-LLM \cite{ado-llm} integrates BO with in-context learning, leveraging LLMs to suggest design points based on prior knowledge.

However, for more efficient analog design, emulating human designers through information reuse is crucial. While \cite{gcn_rl} applies transfer learning through graph-based optimization, it lacks a semantic understanding of the design space. To address this, we propose LLM-USO, which combines the strengths of LLMs and BO to achieve more efficient analog design optimization with enhanced information reuse.

\section{Problem Formulation}

Formally, let $\mathcal{C}_i$ represent circuit $i$ with sub-structures $\mathcal{S}_i = \{S_{i_1}, S_{i_2}, \dots, S_{i_n}\}$, and let $\mathcal{X}_i$ represent the design parameter space and $\mathcal{P}_i$ represent the space of the set of performance metrics for this circuit. The objective function for circuit $i$, denoted by $f_i(\mathbf{x}_i)$, (where $\mathbf{x}_i \in \mathcal{X}_i$)  aggregates multiple performance metrics $(\{P_{i_1}, P_{i_2}, \cdots \} \in \mathcal{P}_i)$ (e.g., gain, bandwidth) into a figure of merit (FOM) to be maximized or minimized. The challenge lies in efficiently exploring the design space  $\mathcal{X}_i$ to find the optimal design parameters $\mathbf{x}_i^*$ that satisfy the performance specifications. 

Our problem formulation introduces \textit{information reuse} by leveraging structured design knowledge from previously optimized circuits that share common sub-structures with the current circuit. Specifically, given a set of circuits \(\{\mathcal{C}_1, \mathcal{C}_2, \dots, \mathcal{C}_k\}\), with corresponding structured knowledge summaries \(\{\mathcal{K}_1, \mathcal{K}_2, \dots, \mathcal{K}_k\}\), the objective is to reuse knowledge \(\{\mathcal{K}_j : j \in \mathcal{J}\}\), where \(\mathcal{J} \subseteq \{1, \dots, k\}\) is the subset of indices such that the circuits \(\mathcal{C}_i\) and \(\mathcal{C}_j\) share similar sub-structures. The optimization process can be formulated as:

\[
\mathbf{x}_i^* = \arg \max_{\mathbf{x}_i \in \mathcal{X}_i} f_i(\mathbf{x}_i \mid \{\mathcal{K}_j : j \in \mathcal{J}\}),
\]

\section{Background}

\subsection{ADO-LLM framework}

Analog Design Optimization with Large Language Models (ADO-LLM) \cite{ado-llm} integrates Bayesian Optimization (BO) with the contextual reasoning capabilities of large language models (LLMs) to automate analog circuit design by leveraging LLMs' in-context learning to generate design points based on circuit definitions and objectives. The framework operates with a shared data buffer, where BO suggests design points using Gaussian Process (GP) models, and LLMs provide complementary suggestions informed by pre-trained circuit knowledge. ADO-LLM enhances BO's efficiency, particularly in high-dimensional design spaces, and accelerates convergence through the integration of broad design knowledge. However, unlike human designers, ADO-LLM cannot reuse information from previous circuit optimizations. Moreover, its exploration is limited by accepting only a single LLM suggestion per iteration, which may lead to repeated suggestions and restrict the discovery of novel design points. While ADO-LLM assumes that LLM suggestions will be close to existing data points, LLMs with sufficient design knowledge can propose more diverse and far-reaching points, improving exploration. \textbf{To address these limitations, we propose LLM-USO, which leverages a structured knowledge representation tailored for analog design and enables transfer learning through human-like design strategies.} 

 \begin{figure*}[!htbp]
  \centering
  \includegraphics[width=0.7\textwidth]{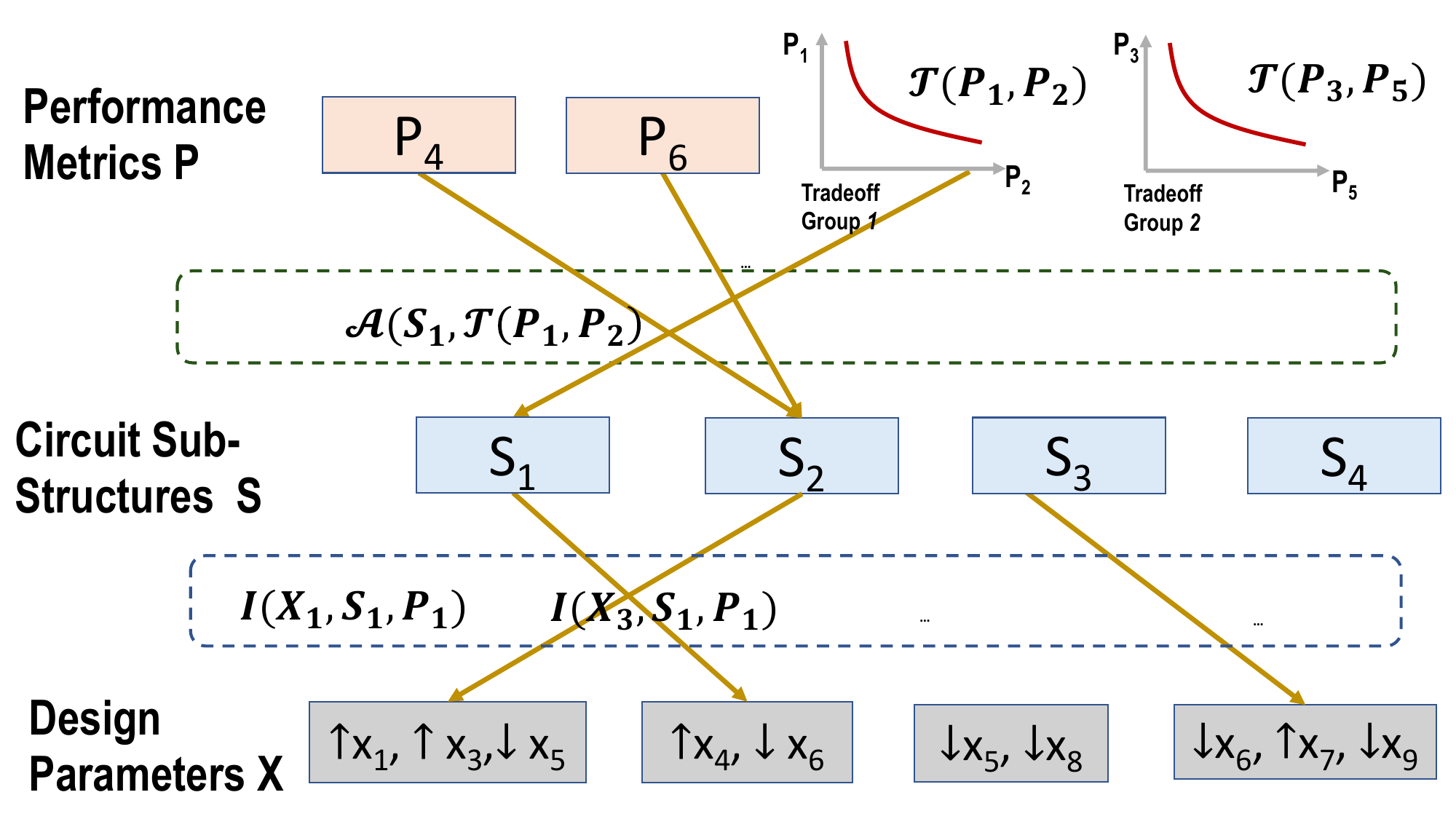}
  \caption{Illustration of analog design  knowledge representation}
  \label{fig:knowledge_representation}
\end{figure*}

\section{Structured Representation of Analog Design Knowledge}

\label{sec:knowledge_representation}

Optimizing analog circuits is inherently challenging due to the complex interplay between performance metrics (\(\mathcal{P}_i\)) and design parameters (\(\mathcal{X}_i\)) within a given circuit (\(\mathcal{C}_i\)). The relationships between these elements encapsulate valuable analog design knowledge. However, fully utilizing this knowledge requires a deeper understanding of how intermediate circuit sub-structures (\(\mathcal{S}_i\)) mediate these interactions. Sub-structures such as differential pairs, current mirrors etc. are commonly shared across different circuits, serving as natural building blocks that group related design parameters and performance metrics. Capturing these finer-grained relationships among performance metrics, sub-structures, and design parameters enables a more organized and transferable representation of analog design knowledge. Motivated by this, our work aims to formalize this interplay to enhance the efficiency and effectiveness of analog circuit design.

This section details a structured methodology to represent analog design knowledge in three steps: (1) identifying relationships between performance metrics, (2) associating performance metrics with relevant sub-structures, and (3) linking design parameters within sub-structures to their influence on performance metrics. Our structured representation depicts the causal flow of the analog design process for a given circuit and is depicted in Figure \ref{fig:knowledge_representation}. %and additionally allows for the reuse of design knowledge across circuits with overlapping sub-structures, supporting transfer optimization.

\subsection{Identifying Relationships Between Performance Metrics}

Performance metrics (\(\mathcal{P}\)) for a given circuit \(\mathcal{C}_i\) often exhibit inherent trade-offs. These trade-offs can be formally captured using a grouping operator \(\mathcal{T}\), which defines a trade-off relationship between two performance metrics:
\begin{equation}
\mathcal{T}(P_l, P_m) \quad \, ; \quad \, P_l, P_m \in \mathcal{P}.    
\end{equation}

where \(P_l\) and \(P_m\) represent the \(l^\text{th}\) and \(m^\text{th}\) performance metrics of circuit \(i\), respectively. With a slight abuse of notation, we drop the subscript \(i\) for circuit $\mathcal{C}_i$ for simplicity. For example, \(\mathcal{T}(P_{\text{gain}}, P_{\text{bandwidth}})\) captures the trade-off between gain and bandwidth, highlighting their competing objectives.

\subsection{Associating Sub-Structures with Performance Metrics}

Circuit sub-structures (\(\mathcal{S} = \{S_1, S_2, \dots\}\)) are closely associated with specific performance metrics or their trade-offs. To formalize this relationship, let \(\mathcal{A}\) denote the association between a sub-structure and a performance metric or performance trade-off pair:
\begin{equation}
\mathcal{A}(S_j, \mathcal{T}(P_l, P_m)) \quad ; \, S_j \in \mathcal{S}, \, P_l, P_m \in \mathcal{P}.    
\end{equation}

Here, \(S_j\) represents the \(j^\text{th}\) sub-structure, and \(P_l\) and \(P_m\) are the \(l^\text{th}\) and \(m^\text{th}\) performance metrics, respectively. With a slight abuse of notation, we drop the subscript \(i\) (for circuit $\mathcal{C}_i$) for simplicity. For example, \(\mathcal{A}(S_1, \mathcal{T}(P_{\text{gain}}, P_{\text{bandwidth}}))\) associates a differential pair sub-structure (\(S_1\)) with the gain-bandwidth trade-off.

\subsection{Linking Design Parameters to Sub-Structures and Performance Metrics}

Design parameters (\(\mathcal{X} = \{X_1, X_2, \dots\}\)) of a circuit are grouped by sub-structures (\(S_j\)) and by their influence on performance metrics (\(P_l\)). To formalize this relationship, let \(\mathcal{I}\) represent the directional influence of a design parameter \(X_p\) within a sub-structure \(S_j\) on a performance metric \(P_l\):

\begin{equation}
\mathcal{I}(X_p, S_j, P_l) \quad ; \, X_p \in \mathcal{X}, \, S_j \in \mathcal{S}, \, P_l \in \mathcal{P}.    
\end{equation}

Here, \(X_p\) refers to the \(p^\text{th}\) design parameter belonging to the sub-structure \(S_j\), and \(P_l\) denotes the \(l^\text{th}\) performance metric influenced by it. With a slight abuse of notation, the subscript \(i\) indicating the circuit is omitted for simplicity.

Our formulation explicitly associates a design parameter with the corresponding sub-structure that it belongs to and highlights how the design parameter impacts the performance metric. For instance, \(\mathcal{I}(X_{W_1}, S_1, P_{\text{gain}})\) captures how the transistor width \(X_{W_1}\) within the sub-structure \(S_1\) (e.g., a differential pair) influences the performance metric \(P_{\text{gain}}\). 

\subsection{Unified Knowledge Representation}

Building on these proposed components, the structured knowledge representation for a circuit \(\mathcal{C}_i\) is defined as:

\begin{multline}
\mathcal{K}_i = \Big\{ \cdots, 
\big(\mathcal{T}(P_l, P_m), 
\mathcal{A}(S_j, \mathcal{T}(P_l, P_m)), \\ 
\mathcal{I}(X_p, S_j, P_l)\big), \cdots \Big\}
\end{multline}

This knowledge representation captures the complex dependencies of performance metrics, or their trade-offs, on design parameters by incorporating intermediate sub-structure information. The fine-grained detail provided by sub-structures enables a deeper understanding of circuit behavior and proves particularly valuable for cross-circuit transfer learning. Since many analog circuits share similar sub-structures, this representation, which encodes interdependencies at the sub-structure level, is well-suited for transfer learning, allowing the effective reuse of knowledge from one circuit to another.

\section{Large Language Model-based Universal Sizing Optimizer : LLM-USO}

Building on the structured knowledge representation defined in Section \ref{sec:knowledge_representation}, we propose the Large Language Model-based Universal Sizing Optimizer (LLM-USO). This hybrid framework combines Bayesian Optimization (BO), large language models (LLMs), and knowledge reuse enabled by the structured representation to efficiently optimize analog circuit designs. By leveraging this unified approach, LLM-USO facilitates transfer learning and enhances the efficiency of the design process. We illustrate the flow of LLM-USO in Figure \ref{fig:LLM-USO_flow} and provide the algorithmic flow in Algorithm \ref{alg:llm-uso}. % Our framework operates in two distinct stages.

Leveraging the structured knowledge summaries (\(\mathcal{K}_1, \ldots, \mathcal{K}_j\)) of related circuits with similar sub-structures, LLM-USO employs a hybrid framework combining BO and LLMs to optimize the current circuit \(\mathcal{C}_i\). For each circuit to be optimized, LLM-USO initializes a data buffer with initial design points and iteratively refines its understanding of the design space.

At each iteration \(t\), the framework performs two key tasks. First, it utilizes the BO-LLM framework to recommend a design point \(\mathbf{x}^t\) by combining the context provided by knowledge summaries (\(\mathcal{K}_1, \ldots, \mathcal{K}_j\)) of related circuits. Second, it updates the data buffer with the observed performance of \(\mathbf{x}^t\) and dynamically generates a knowledge summary \(\mathcal{K}_i\) for the current circuit using the circuit context, previously generated examples, and structured prompts.

Upon completing the iterations, LLM-USO also performs a final refinement of the knowledge summary \(\mathcal{K}_i\) through a critiquing process to ensure its quality and relevance. The framework then outputs both the optimal design parameter \(\mathbf{x}^*\) and the refined knowledge summary \(\mathcal{K}_i\), enabling efficient optimization of the current circuit while contributing to reusable design knowledge for future circuits.

In this section, we first present the practical methodology for generating knowledge summaries (Section \ref{subsec:KS_generation}), followed by the introduction of the knowledge critiquing mechanism (Section \ref{subsec:K_critiquing}). Next, we describe how knowledge summaries from related circuits are incorporated into the LLM's context to facilitate transfer learning (Section \ref{subsec:information_reuse_context}). Finally, we propose a novel exploration mechanism within the knowledge reuse-based hybrid BO-LLM framework to enable efficient optimization (Section \ref{subsec:UCB_ranking}).

\begin{algorithm}[!htbp]
\caption{LLM-USO Framework}
\label{alg:llm-uso}
\noindent \textbf{Input:} Knowledge summary of \textbf{related} circuits \( \{\mathcal{K}_1, \ldots, \mathcal{K}_{j}\} \)\\
\textbf{Output:} Optimal design parameter of current circuit $C_i$ - \( (\mathbf{x}^*) \), knowledge summary of current circuit $\mathcal{K}_i$ \

Initialize common data buffer to $B = \{\}$\

Collect $I$ initial design points and populate buffer $B$. 

\begin{enumerate}[leftmargin=*, label=\arabic*.]
    \item For $t \in [1, T]$ :
    \begin{enumerate}[leftmargin=*, label=\alph*.]
         \item Recommend design point $\mathbf{x}^t $ from the BO-LLM framework using the information reuse context \( \{\mathcal{K}_1, \ldots, \mathcal{K}_{j}\} \) and Uncertainty based ranking of the LLM recommendations $\triangleright$ \textcolor{red}{(Sec VI-C and VI-D)} 
        \item  Update buffer $B$ with  \( (\mathbf{x}^t, f(\mathbf{x}^t) ) \)
        \item Generate knowledge summary $\mathcal{K}_i$ using circuit context, top demonstration examples from $B$ and  structured prompts. $\triangleright$ \textcolor{red}{(Sec VI-A)}

    \end{enumerate}

Refine $\mathcal{K}_i$ through Knowledge critiquing $\triangleright$ \textcolor{red}{(Sec VI-B)} \

Output $\mathbf{x^*}, \mathcal{K}_i$
\end{enumerate}
\end{algorithm}
%\vspace{-1em}

\subsection{Structured Design Knowledge Summary Generation}
\label{subsec:KS_generation}

\begin{figure*}[!htbp]
  \centering
  \includegraphics[width=0.75\textwidth]{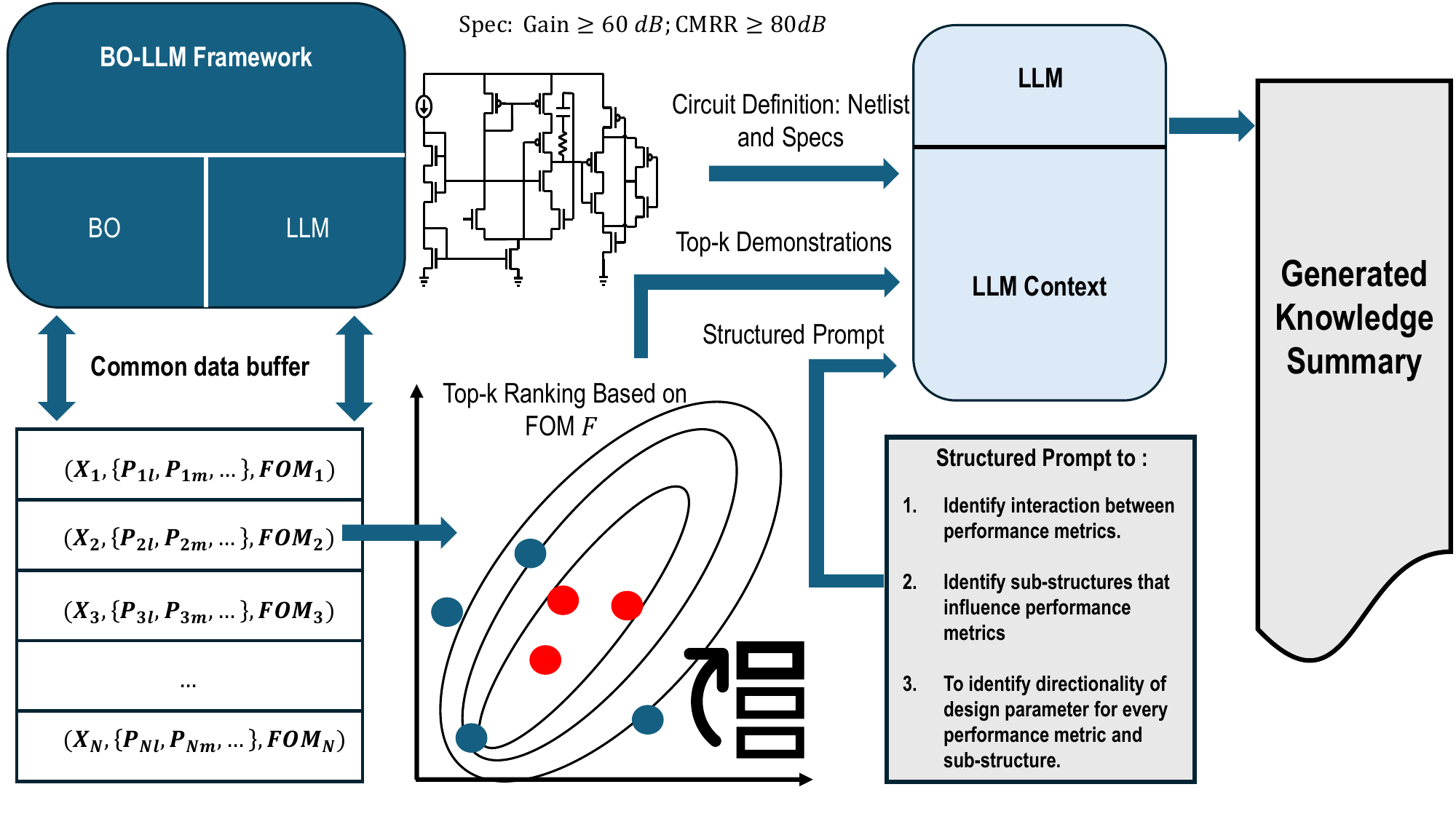}
  \caption{Knowledge Summary Generation: Involves using the circuit definition, top demonstration examples in the common data buffer and a structured prompt design as inputs to the LLM top generate the structured design knowledge summary. Top demonstration examples are sampled from the common data buffer using top-K sampling. }
  \label{fig:netlist_fewshot}
\end{figure*}

A structured knowledge summary is a text-based representation of the design rules outlined in Section \ref{sec:knowledge_representation} (and illustrated in Figure \ref{fig:knowledge_representation}). In essence, it encapsulates the sizing details of design parameters associated with various circuit sub-structures and establishes their connections to the resulting performance metrics or performance trade-off pairs.

To practically generate such comprehensive knowledge representations at each iteration $t$ of the optimization process,  LLM-USO employs three core components. It uses the netlist and design specifications as circuit definition, top demonstration examples and a novel structured prompt design as inputs to an LLM to generate the structured knowledge summary. We illustrate this in Figure \ref{fig:netlist_fewshot}. First, to encode general design rules based on the circuit’s structure, it uses the circuit definition, represented by the netlist, along with design objectives, as inputs to the LLM. Second, to enable the model to infer how relative sizing affects the performance of substructures and, ultimately, the design objectives, it leverages the best design points obtained from the LLM-BO optimization (such as from ADO-LLM) as top demonstration examples. Lastly, to generate the intended structured design knowledge as in Section \ref{sec:knowledge_representation}, LLM-USO employs a novel structured prompt design. This prompt guides the LLM to extract causal relationships that show how changes in design points belonging to particular sub-structures affect the overall design objectives while considering any trade-offs when applicable.

\subsubsection{\textbf{Circuit Definition as Context}}
To explain the first part of the structured design knowledge summary generation in LLM-USO, we begin by leveraging the circuit definition and design objectives as key inputs to the LLM. This process is similar to ADO-LLM, where the netlist—a detailed representation of the circuit’s topology, serves as the foundational input.

For a given circuit, the structured representation of the netlist provides the LLM with a detailed understanding of the circuit’s architecture. In LLM-USO, the netlist is combined with specific design objectives which guide the optimization process. These objectives inform the LLM of the target performance metrics that the design aims to achieve. By presenting both the netlist and the design objectives as context, the LLM is able to reason about the circuit at both the structural and functional levels. This allows the model to infer general design rules that are applicable to the circuit being optimized. The LLM processes this information in a structured prompt, where the netlist and design objectives are represented textually in a format that the model can interpret/

\subsubsection{\textbf{Top Demonstration Examples as Context}}

In the second step of LLM-USO’s structured design knowledge summary generation process, we leverage top demonstration examples (key design points obtained in prior optimization runs of the same circuit) to provide the LLM with concrete examples of how the sizing of different circuit sub-structures affects the overall performance. These top demonstration examples are obtained through the combined interaction of Bayesian Optimization (BO) and LLM-guided exploration, leveraging the best simulation results to extract meaningful interactions between the design points and final performance metrics.

Specifically, the combined design point recommendation by both the BO and LLM agents in LLM-USO (similar to ADO-LLM \cite{ado-llm}), is stored in a common data buffer. The data buffer acts as a repository of all the simulated design points, their corresponding performance metrics and figure of merit (FOM) values. From the shared buffer, the top-performing design points are selected as top demonstration examples based on a ranking criterion, such as the top-k points by Figure of Merit (FOM), which combines multiple performance metrics. %(In practice, we use top-3 sampling).
These examples highlight critical design rules and sub-structure relationships. To finally ellicit a suitable design knowledge summary, the selected top demonstration examples are formatted into a text format that the LLM can interpret. This format includes:
\begin{itemize}
    \item Parameter values for key sub-structures (e.g., transistor dimensions, bias currents).
    \item Corresponding performance metrics.
\end{itemize}
By incorporating these examples into the LLM's prompt among other components, the model learns to identify the causal correspondence between the sizing of the sub-structures and the performance metrics. We illustrate the sampling of the top demonstration examples from the buffer in Figure \ref{fig:netlist_fewshot}.

 \begin{figure*}[!htbp]
  \centering
  \includegraphics[width=0.8\textwidth]{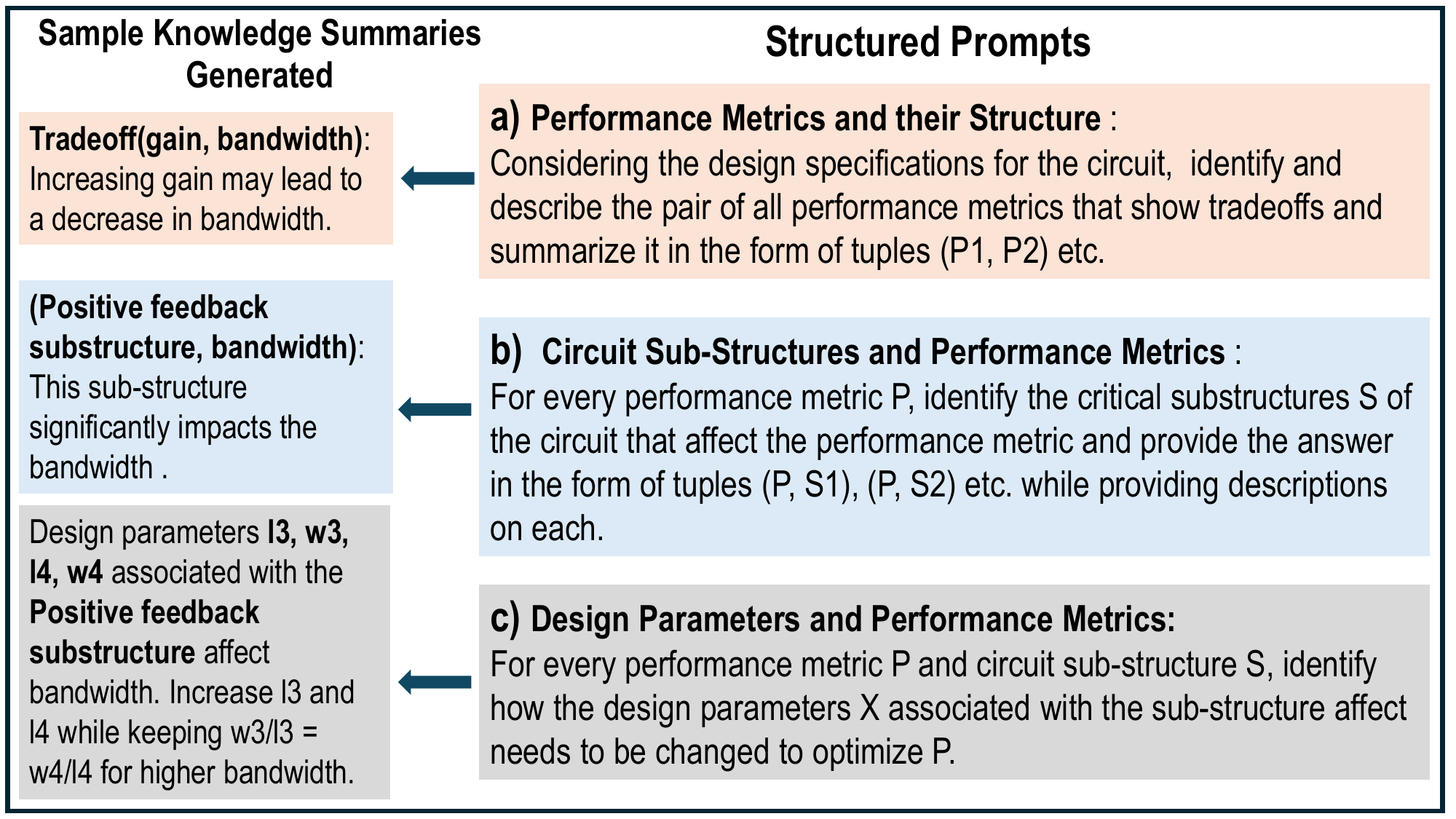}
  \caption{Illustration of the prompt design to extract structured design knowledge}
  \label{fig:structured_prompt}
\end{figure*}

\subsubsection{\textbf{Prompt Design to Extract Structured Design Knowledge Summary}}

Building on the knowledge representation framework outlined in Section \ref{sec:knowledge_representation}, we adopt a systematic and structured prompt engineering approach to extract design knowledge from a circuit.  Given the circuit context and top demonstration examples, this process involves querying the LLM with targeted prompts to extract insights in three critical areas:

\begin{enumerate}
    \item \textit{Linking Performance Metrics and Their Trade-offs}:  
    The LLM is prompted to identify trade-offs between performance metrics (\(P\)) by summarizing pairs of metrics that exhibit competing objectives. 
    
    The prompt requests tuples of the form \((P_1, P_2)\), capturing relationships that help contextualize how optimizing one performance metric impacts others. An illustration of the prompt and the resulting summary is show in Figure \ref{fig:structured_prompt}a.

    \item \textit{Linking Circuit Sub-Structures and Performance Metrics}:  
    The LLM is then guided to associate every performance metric $P$ with the critical sub-structures (\(S\)) in the circuit that directly influence them. 
    This step provides insights into which sub-structures of the circuit are most relevant to optimizing specific performance objectives. An illustrative prompt and the resulting summary are shown in Figure \ref{fig:structured_prompt}b.

    \item \textit{Linking Design Parameters and Performance Metrics via Sub-Structures}:  
    Finally, to identify how design parameters (\(X\)) within specific sub-structures (\(S\)) affect the performance metrics (\(P\)), the LLM  is prompted to provide a detailed mapping of the directional influence of design parameters on performance metrics, enabling fine-grained optimization. Sample prompt and the resulting summary are shown in Figure \ref{fig:structured_prompt}c.
\end{enumerate}

These structured prompts ensure that the LLM generates a comprehensive knowledge summary that captures the dependencies between performance metrics, sub-structures, and design parameters.

\subsection{Knowledge Critiquing Framework}
\label{subsec:K_critiquing}

While LLM-USO generates structured design knowledge summaries, as described in the previous subsection, it is essential to assess the quality and completeness of these summaries. In the LLM-USO framework, a cost-effective and fast working LLM (GPT-3.5) is employed to generate the knowledge summary based on the circuit context, top demonstration examples, and structured prompts at every optimization run. However, since these working LLMs may sometimes produce generic or incomplete knowledge summaries, it becomes necessary to critically evaluate and refine the generated content to ensure it fully captures the nuances of the circuit design process.

\begin{figure*}[!htbp]
  \centering
  \includegraphics[width=0.8\textwidth]{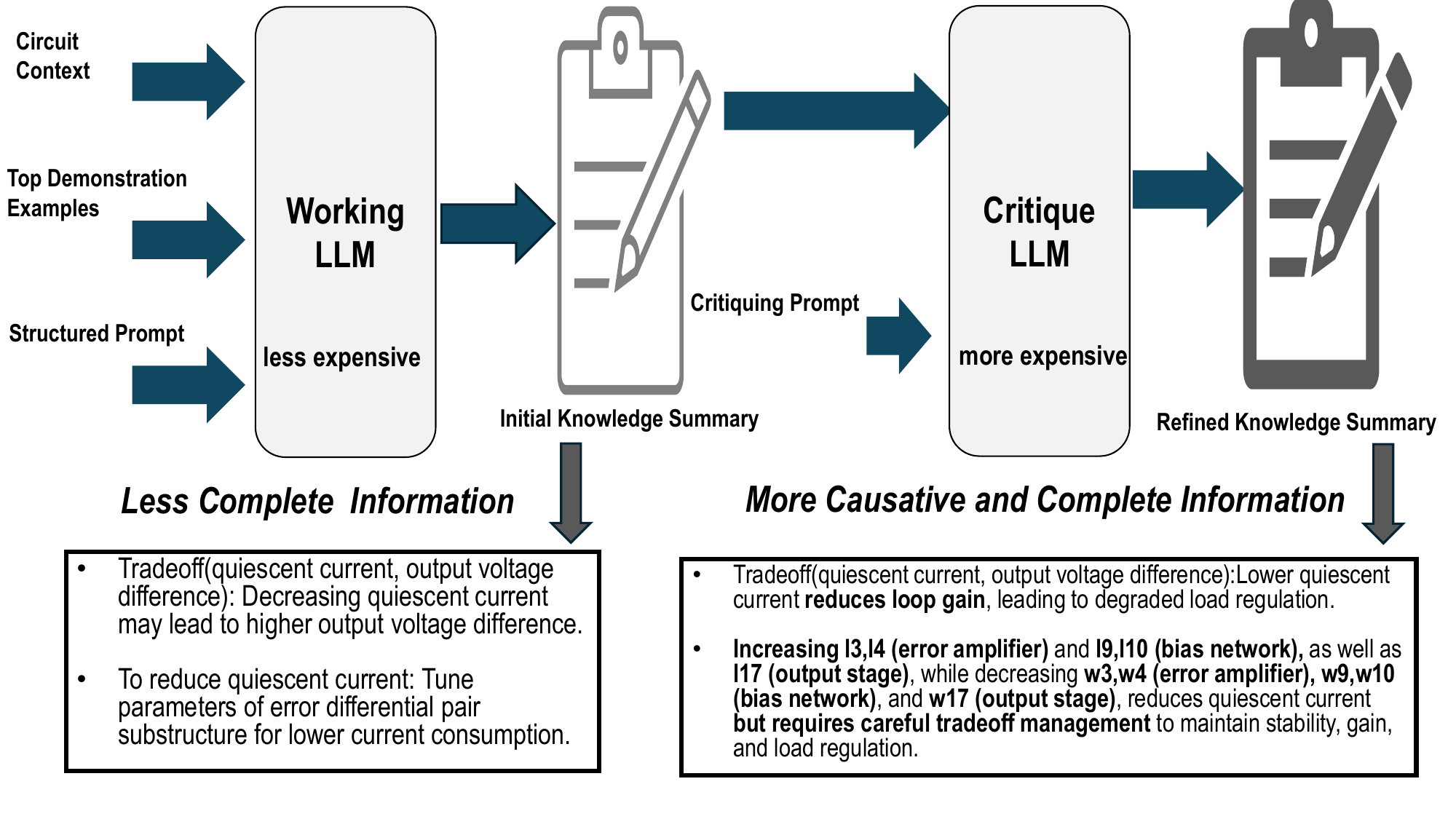}
  \caption{Illustration of the refined knowledge summary generation for sample statements from the low dropout regulator. The refined summary consists of extra context relevant information which can help in transfer learning.}
  \label{fig:critique_framework}
\end{figure*}

To address this, we integrate a more advanced and domain-informed critique LLM (GPT-4) to review and enhance these knowledge summaries. We illustrate our critiquing framework in Figure \ref{fig:critique_framework} (and also show sample corrections suggested in the refined knowledge summary for the low dropout regulator circuit - Figure \ref{fig:ckt_schematics}a). The critique LLM is provided with the initially generated knowledge summary alongside a critiquing prompt that instructs it to identify inconsistencies in the initial knowledge summary and suggest corrections. This process outputs a refined knowledge summary with more precise and comprehensive design insights. As illustrated in Figure \ref{fig:critique_framework}, the critiquing process enhances the original summary by incorporating circuit-specific details (e.g., recognizing why low quiescent current in the low dropout regulator leads to high output voltage difference, and pointing out more complete details on how to reduce quiescent current), resulting in more relevant and complete design knowledge summarization.

A natural question arises: why not use the critique LLM directly for generating the knowledge summaries? The answer lies in the iterative nature of the optimization process, where the knowledge summary (for the concerned circuit) evolves at every iteration due to updates in top demonstration examples. Relying on the high-cost critique LLM for each iteration would be prohibitive. Instead, the low-cost working LLM is used efficiently within the optimization loop to generate knowledge summaries at each iteration. At the end of the optimization process, a single critique is performed using the high-cost critique LLM, producing a refined knowledge summary with the most accurate and targeted insights about the circuit.

This approach effectively balances computational efficiency and the demand for high-quality knowledge summaries. By combining the strengths of both working and critique LLMs, we ensure that the refined summaries are not only reliable but also cost-effective, making them ideal for reuse in subsequent optimization tasks.

\subsection{Information Reuse Context}
\label{subsec:information_reuse_context}
LLM-USO leverages the reuse of the generated high quality structured knowledge summary from previously optimized circuits to enhance the optimization efficiency of the current circuit. Structured design knowledge, along with the corresponding netlists of similar circuits, is provided as additional context to the LLM during its in-context learning phase \cite{in_context_learning}. By incorporating this relevant information, the LLM can apply insights from circuits with similar sub-structures, allowing it to identify optimal design points more effectively. This approach mimics the behavior of expert human designers, who intuitively reuse knowledge from prior designs of similar circuits, to accelerate the optimization of the current circuit.% By reusing high-quality knowledge, LLM-USO reduces computational overhead while improving convergence speed toward optimal solutions.

\subsection{Uncertainty Based Ranking of LLM Suggestions}
\label{subsec:UCB_ranking}

In the LLM-USO framework, both BO and the LLM, informed by structured design knowledge from related circuits, suggest new design points. However, requesting a single suggestion from the LLM may result in repetitive design points, leading to stagnation in exploring novel and promising regions of the design space. While the LLM can generate contextually relevant suggestions, it lacks the ability to identify if a design point has already been explored during the optimization process. On the other hand, requesting multiple design points from the LLM increases the number of costly simulations.

\begin{figure}[!htbp]
  \centering
  \includegraphics[width=0.5\textwidth]{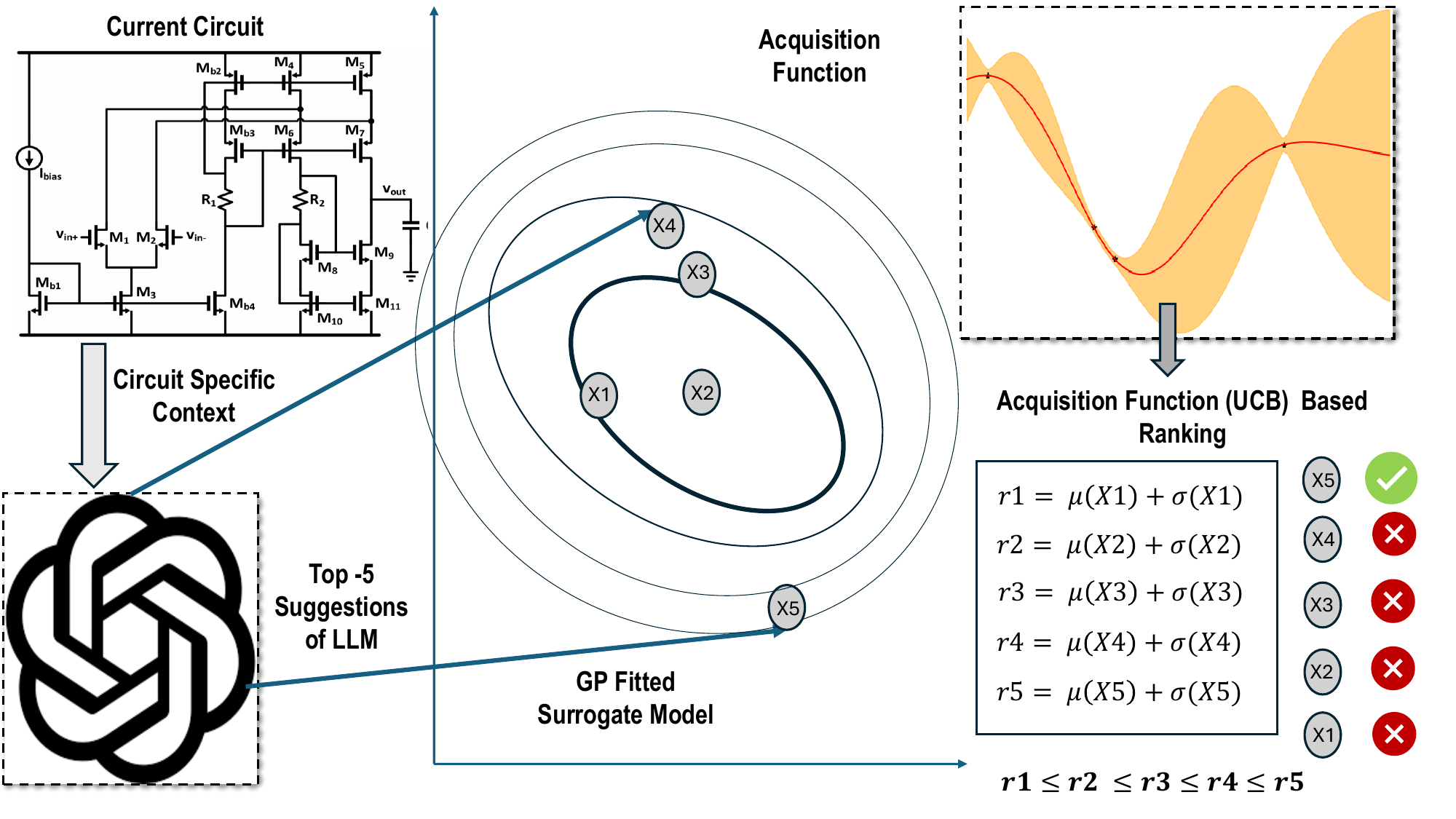}
  \caption{Uncertainty Based Ranking of the LLM Suggestions}
  \label{fig:ucb_ranking}
\end{figure}

To address this, we leverage the uncertainty estimates provided by the Gaussian Process (GP) surrogate model in BO to rank different suggestions by the LLM. We illustrate this in Figure \ref{fig:ucb_ranking}. The GP surrogate models the performance landscape by predicting the mean performance (\(\mu(x)\)) and the uncertainty (\(\sigma(x)\)) associated with each design point \(x\). These predictions enable the computation of the upper confidence bound (UCB) acquisition function \cite{BO_book}, which prioritizes novel and promising regions of the design space. The UCB is defined as:
\[
    \mathrm{UCB}(x) = \mu(x) + \kappa \sigma(x),
\]
where \(\mu(x)\) represents the predicted mean performance of design point \(x\), \(\sigma(x)\) is the standard deviation (uncertainty) of the prediction, and \(\kappa\) is a tunable parameter that balances exploration and exploitation. In all experiments, we set \(\kappa = 1.0\).

We propose to use this BO based UCB acquisition function, to rank the LLM-suggested design points based on their potential to balance high predicted performance and high uncertainty, pointing to unexplored yet valuable regions of the design space. For each batch of suggestions made by the LLM, we rank the design points according to their UCB values. This ranking ensures that the optimization process prioritizes design points that are novel and likely to lead to significant performance improvements. By simulating only the top-ranked design point among many LLM suggested design points, we avoid redundancy and stagnation, ensuring efficiency in the optimization process.

This synergy between the GP surrogate model, BO, and design knowledge enables faster convergence toward optimal designs for a given circuit.

\section{Experiments}

\begin{figure*}[!htbp]
    \centering
    \begin{subfigure}[b]{0.495\textwidth}
        \centering
        \includegraphics[width=\textwidth]{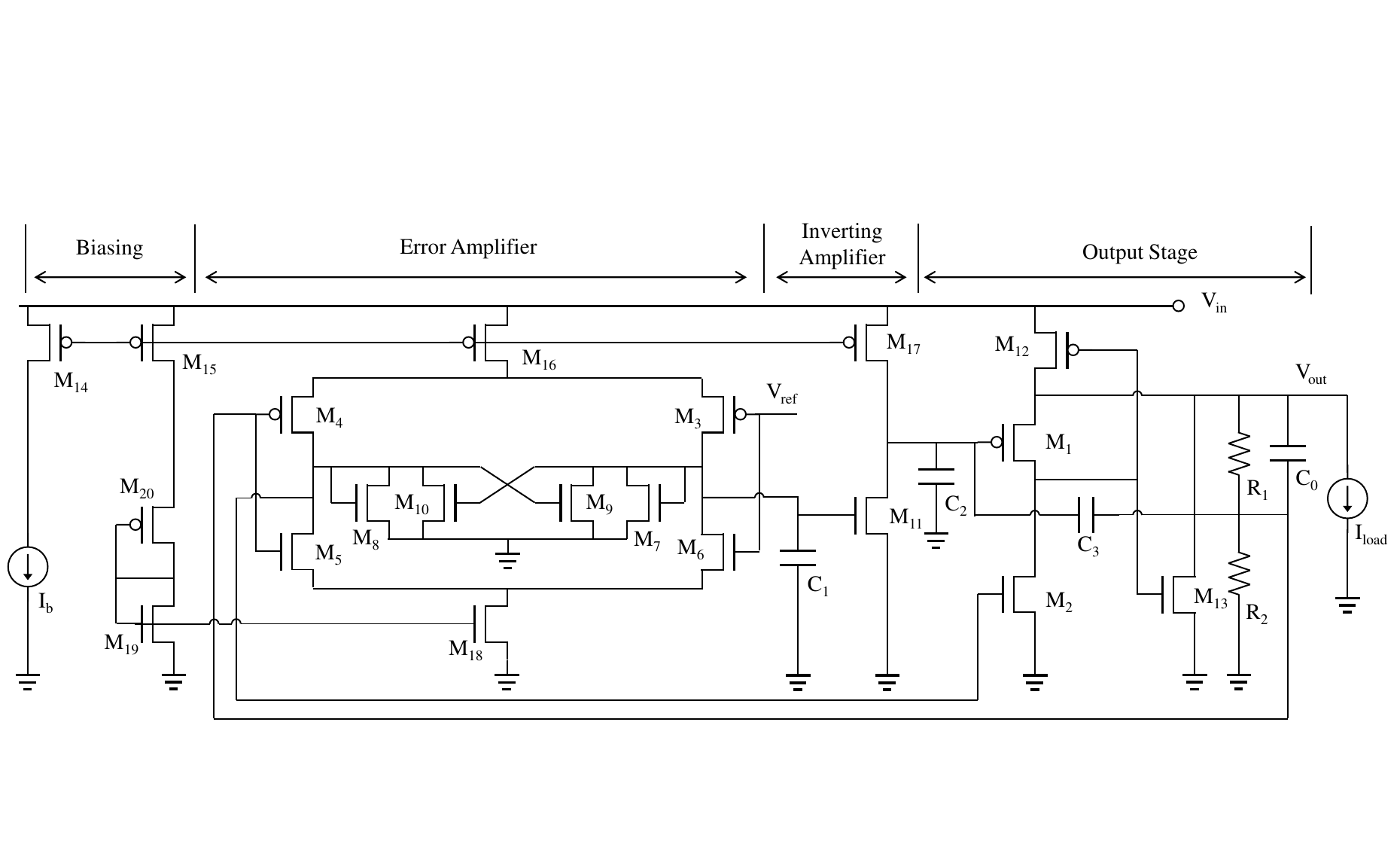}
        \caption{Low Dropout Regulator}
        \label{fig:sub4}
    \end{subfigure}
    %\hfill
    \begin{subfigure}[b]{0.495\textwidth}
        \centering
        \includegraphics[width=\textwidth]{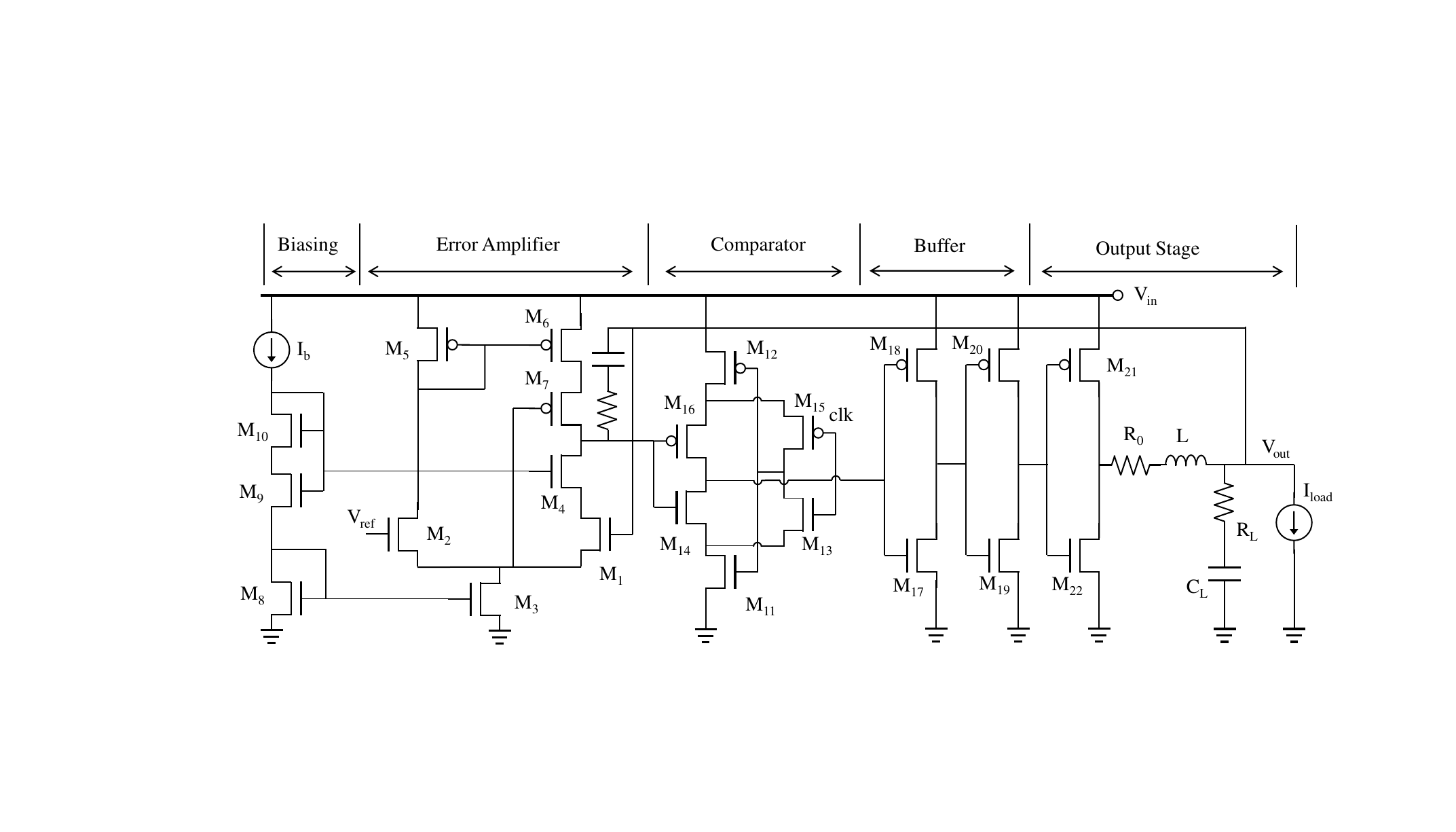}
        \caption{DCDC Converter}
        \label{fig:sub5}
    \end{subfigure}
    
    \begin{subfigure}[b]{0.3\textwidth}
        \centering
        \includegraphics[width=\textwidth]{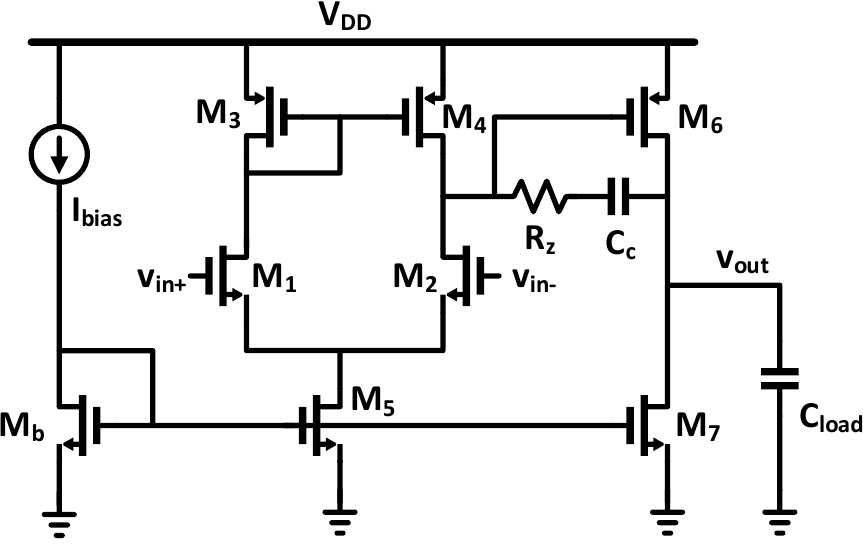}
        \caption{Two Stage Differential Amplifier}
        \label{fig:sub1}
    \end{subfigure}
    %\hfill
    \begin{subfigure}[b]{0.3\textwidth}
        \centering
        \includegraphics[width=\textwidth]{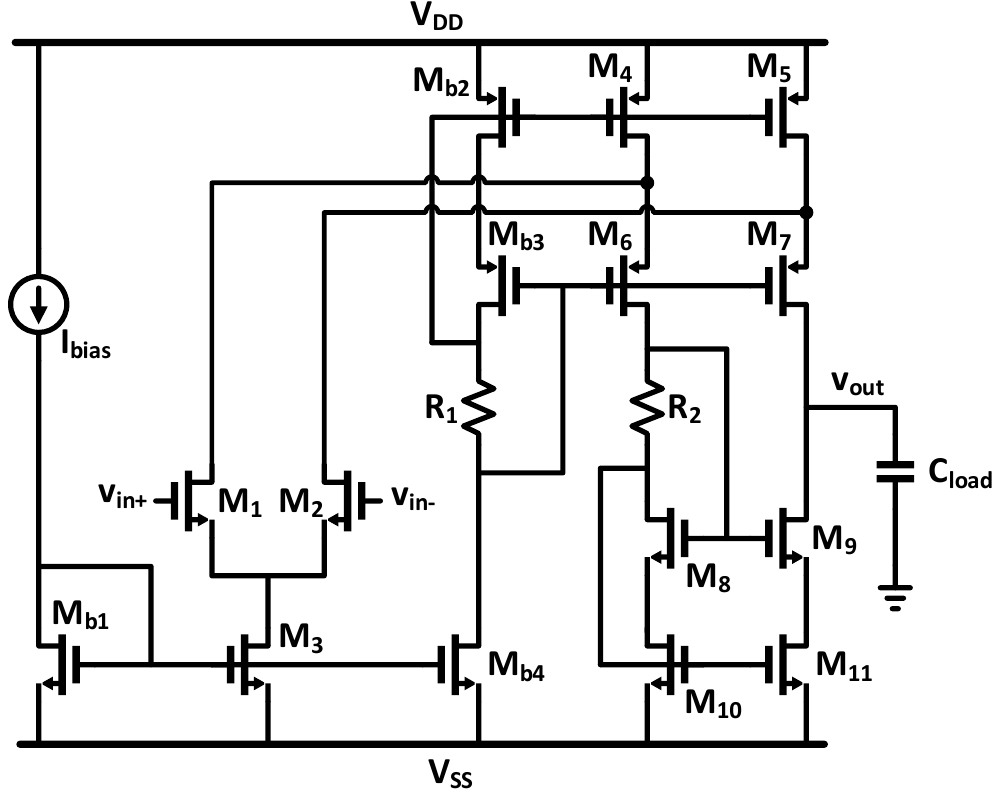}
        \caption{Folded Cascode Amplifier}
        \label{fig:sub2}
    \end{subfigure}
    %\hfill
    \begin{subfigure}[b]{0.3\textwidth}
        \centering
        \includegraphics[width=\textwidth]{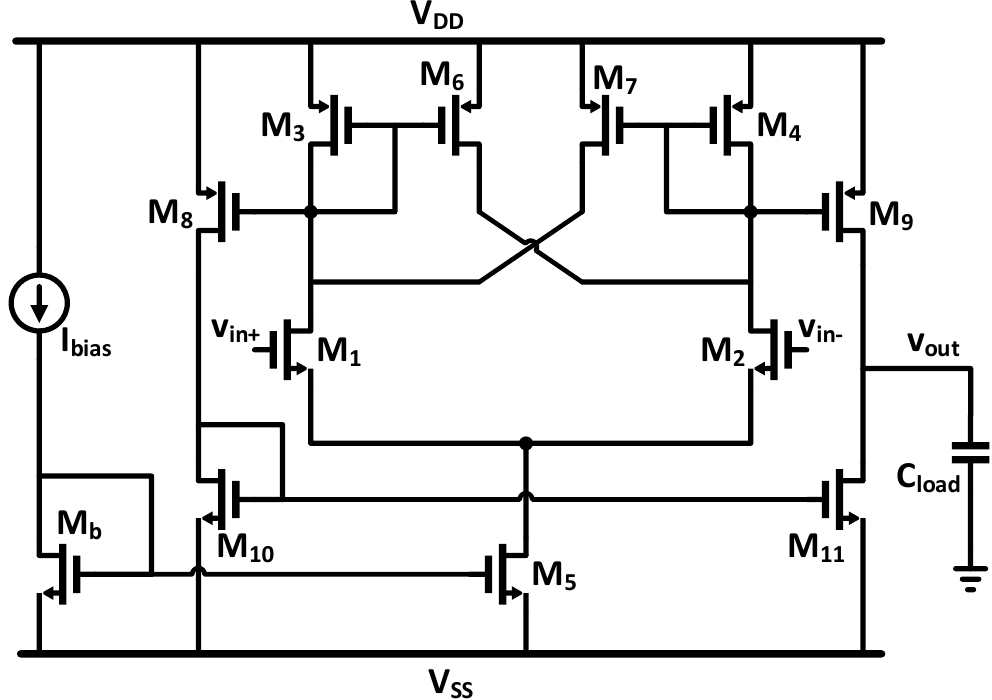}
        \caption{Hysteresis Comparator}
        \label{fig:sub3}
    \end{subfigure}
     %\vspace{-0.5em} % Space between rows

    \caption{Circuit Schematics.}
    \label{fig:ckt_schematics}
    \vspace{-1em}
\end{figure*}

To evaluate the performance of the proposed LLM-USO framework, we conducted experiments on a diverse set of analog circuits, including the two-stage differential amplifier (Figure \ref{fig:ckt_schematics}c), folded cascode amplifier (Figure \ref{fig:ckt_schematics}d), hysteresis comparator (Figure \ref{fig:ckt_schematics}e), low dropout regulator (LDO) (Figure \ref{fig:ckt_schematics}a), and DC-DC converter (DCDC) (Figure \ref{fig:ckt_schematics}b). These circuits, which share several similar sub-structures (with mild change in topology), provide an ideal testbed for demonstrating the benefits of information reuse in the optimization process. We systematically experiment with two variants of LLM-USO to isolate and highlight the benefits of (i) information reuse and (ii) uncertainty-based ranking of LLM suggestions. The first variant, LLM-USO(R), focuses solely on information reuse through structured design knowledge summaries, while the second variant, LLM-USO(C), combines information reuse with the integration of uncertainty-based ranking. In LLM-USO(C), the uncertainty metric is used to rank LLM-generated design points based on the upper confidence bound (UCB) of the BO surrogate model.

To assess the performance of LLM-USO(R) and LLM-USO(C), we compare them against the baseline ADO-LLM framework \cite{ado-llm}, demonstrating how the injection of relevant information from previous circuit optimizations can improve current optimization processes. Additionally, we compare the LLM-USO variants with a standard GP-BO approach \cite{BO}. We provide the details of our setup in Tables \ref{tab:bo_llm_interaction}, \ref{tab:knowledge_summary_creation}, \ref{tab:sim_budget}. As in ADO-LLM, we start each optimization process with 5 initial design points suggested by the LLM agent. In both LLM-USO(R) and ADO-LLM, at each iteration, the LLM agent proposes 1 candidate design point, while the BO model suggests 1 additional point for simulation evaluation. In contrast, for LLM-USO(C), the LLM agent proposes 4 candidate design points, which are ranked based on the UCB acquisition function. The top-ranked LLM design point, along with the 1 design point suggested by the BO model, is evaluated through simulation.

To maintain consistency, we limit the optimization process to 20 iterations for all methods as shown in Table \ref{tab:sim_budget}. As a result, in GP-BO, ADO-LLM, LLM-USO(R), and LLM-USO(C), the total number of simulation evaluations is set to $5 + 2 \times 20 = 45$. This stricter simulation budget, compared to ADO-LLM, is used to assess whether information reuse can achieve design specifications with fewer simulations. As per the LLM in the BO-LLM interaction setup, we employed GPT-3.5 with a temperature of 0.5. For the working-LLM to generate a knowledge summary we use GPT-3.5 and employ GPT-4 for the critique-LLM. These details are shown in Table \ref{tab:knowledge_summary_creation}. For the BO in LLM-USO and ADO-LLM, we employ the qEI acquisition function \cite{qEI} and a RBF kernel. We define the FOM for a given circuit $C_i$ as $FOM_i =  \sum_j (-1)^{s_j}\Tilde{p_j}$, where $s_j$ is 0 if the metric needs to be maximized and 1 otherwise. $\Tilde{p_j}$ is the normalized performance metric and $j$ is the index of the performance metric.

%\subsection{Two-Stage Differential Amplifier}
We begin our experiments with the two-stage differential amplifier, which serves as a foundational analog circuit for testing our framework. Since this is the initial circuit in our evaluation, we do not have previous design knowledge summaries to leverage for information reuse. Therefore, we generate a structured knowledge summary for the two stage differential amplifier with ADO-LLM as a baseline. The objective in optimizing the two-stage differential amplifier is to achieve a gain greater than 60 dB, a common-mode rejection ratio (CMRR) greater than 45 dB, and a unity-gain frequency (UGF) exceeding 1 MHz.

To illustrate the transfer learning process for subsequent circuits, Table \ref{tab:circuit_context} highlights the circuits used in the combined context in LLM-USO, based on similar sub-structures they share.

\begin{table}[h!]
\centering
\caption{Set-up of the hybrid BO-LLM interaction in compared methods}
\begin{tabular}{@{}lll@{}}
\toprule
\textbf{Sub-model} & \textbf{Hyper-parameter}              & \textbf{Value} \\ \midrule
LLM agent          & \# Queries per Step  in ADO-LLM, LLM-USO(R)               & 1              \\
 & \# Queries per Step  in LLM-USO(C)               & 4              \\
                   & \# Top demonstration examples          & 3             \\
                   & LLM Version                          & GPT 3.5        \\
                   & Temperature                          & 0.5            \\
                   & Context Window Length                & 16k            \\
                   & Max Token Generation Length          & 1,000          \\ \midrule
BO              & \# Query per Step                    & 1              \\
                   & Kernel Function                      & RBF            \\
                   & Acquisition                          & qEI {[}3{]}    \\
                   & Acquisition Optimizer                & L-BFGS         \\ \bottomrule
\end{tabular}
\label{tab:bo_llm_interaction}
\end{table}

\begin{table}[h!]
\centering
\caption{Set up of knowledge summary generation mechanism}
\begin{tabular}{@{}lll@{}}
\toprule
\textbf{Sub-model} & \textbf{Hyper-parameter}              & \textbf{Value} \\ \midrule
Working-LLM        
                   & \# Top Examples in Context for summary creation          & 3             \\
                   & LLM Version                          & GPT 3.5        \\
                   & Temperature                          & 0.5            \\
                   & Context Window Length                & 16k            \\
 \midrule
critique-LLM              & LLM Version                          & GPT 4        \\
& Temperature          & 0.5            \\
& Context Window Length                & 16k            \\
\bottomrule
\end{tabular}
\label{tab:knowledge_summary_creation}
\end{table}

\begin{table}[h!]
\centering
\caption{Details on simulation budget}
\begin{tabular}{@{}lll@{}}
\toprule
 \textbf{Quantity}              & \textbf{Value} \\ \midrule
 \# Initial data samples $(I)$          & 5             \\
 \# Iterations in optimization process $(T)$          & 20             \\
 \# Simulations per optimization time step $(m)$          & 2             \\
 \# Total number of simulations = $I + m \times T$          &      45   \\
  %Exploration paramter $\kappa$ in UCB ranking              & 1.0            \\

\bottomrule
\end{tabular}
\label{tab:sim_budget}
\end{table}

\begin{table*}[!htbp]
\centering
\renewcommand{\arraystretch}{1.5} % Increase row height for better readability
\setlength{\tabcolsep}{6pt} % Adjust column padding
\begin{tabular}{|c|c|c|}
    \hline
    \textbf{Current Circuit} & \textbf{Circuits Used for Context} & \textbf{Similar Substructures} \\
    \hline
    Folded Cascode Amplifier & Two-Stage Differential Amplifier & Differential Pair \\
    \hline
    Hysteresis Comparator & Two-Stage Differential Amplifier & Differential Stage (Cross-Connected) \\
    \hline
    LDO & Two-Stage Differential Amplifier, Hysteresis Comparator & Error Amplifier, Bias Stage \\
    \hline
    DCDC Converter & Two-Stage Differential Amplifier, Hysteresis Comparator, LDO & Error Amplifier, Comparator \\
    \hline
\end{tabular}
\caption{Table describing the relevant circuits used for context enhancement during transfer learning along with corresponding similar sub-structures.}
\label{tab:circuit_context}
\end{table*}

\subsection{Folded Cascode Amplifier}

In the second experiment, we evaluate the performance of LLM-USO on the folded cascode amplifier, which shares commonality in both structure  (like differential pair) and functionality to the two stage differential amplifier. This circuit has 32 design parameters including transistor sizing parameters and resistor values. For this circuit, we compare both variants of our approach, LLM-USO(R) and LLM-USO(C) (where structured knowledge summary of the two stage differential amplifier are given as context to the LLM-USO), to assess the impact of information reuse and the combination of reuse with uncertainty-based ranking. The objective for optimizing the folded cascode amplifier is: achieving a gain greater than 60 dB, a common-mode rejection ratio (CMRR) greater than 80 dB, and a unity-gain frequency (UGF) exceeding 1 MHz. 

\begin{table}[!htbp]
\centering
\scriptsize
\resizebox{\columnwidth}{!}{%
\begin{tabular}{|c|c|c|c|c|}
\hline
\textbf{Method} & \textbf{Gain (dB) $\uparrow$} & \textbf{CMRR (dB) $\uparrow$} & \textbf{UGF (MHz) $\uparrow$} & \textbf{FOM $\uparrow$} \\
\hline
\textbf{Spec Constraint} & $\geq 60$dB & $\geq 80 $dB & $\geq 1$MHz &  \\
\hline
BO \cite{BO} & 38.51 & 45.33 & \textbf{7.61} & 0.307 \\
ADO-LLM \cite{ado-llm} & 42.23 & 87.27 & 1.65 & 0.299 \\
LLM-USO(R) & 56.99 & \textbf{88.24} & 1.57 & 0.419 \\
LLM-USO(C) & \textbf{63.00} & 85.18 & 2.18 & \textbf{0.447} \\
\hline
\end{tabular}
}
\caption{Performance metrics for the folded cascode amplifier. Arrows indicate the desired direction of optimization, and the specification constraints are listed for each metric.}
\label{tab:folded_cascode}
\end{table}

As seen in Table \ref{tab:folded_cascode}, LLM-USO(R) leverages structured design knowledge to improve gain, CMRR, and UGF compared to ADO-LLM by effectively identifying sub-structures and performance trade-offs from the two stage differential amplifier. LLM-USO(C) further improves performance by using BO’s uncertainty to rank LLM suggestions, focusing on high-value design points. In contrast, BO maximizes UGF at the expense of gain and CMRR, illustrating the limitation of relying solely on figure-of-merit (FOM) without domain-specific context. 

\subsection{Hysteresis Comparator}

We next evaluate the Hysteresis comparator, which, despite having a different functionality from the two-stage differential amplifier, shares similar sub-structures such as the cross-connected differential pair and bias stage. This circuit has 24 sizing parameters across its 12 transistors. The target specifications for the hysteresis comparator are a gain greater than 40 dB, bandwidth (B/W) greater than 1 MHz, hysteresis error magnitude less than 70 mV, and offset voltage below 10 mV. From Table \ref{tab:hyst_comp}, LLM-USO(R) leverages prior context from the amplifier to suggest design points. However, its single-point suggestions may sometimes miss certain specifications. In contrast, LLM-USO(C), which ranks multiple LLM-generated suggestions based on BO’s uncertainty, is more effective in finding design points that satisfy nearly all performance metrics. This highlights the importance of uncertainty-driven exploration when using LLM suggestions to navigate the design space effectively.

\begin{table}[ht]
\centering
\scriptsize
\resizebox{\columnwidth}{!}{%
\begin{tabular}{|c|c|c|c|c|c|}
    \hline
    \makecell{\textbf{Method}} & 
    \makecell{\textbf{Gain}\\\textbf{(dB) $\uparrow$}} & 
    \makecell{\textbf{B/W}\\\textbf{(MHz) $\uparrow$}} & 
    \makecell{\textbf{Offset}\\\textbf{(mV) $\downarrow$}} & 
    \makecell{\textbf{Hyst. Err}\\\textbf{(V) $\downarrow$}} & 
    \makecell{\textbf{FOM $\uparrow$}} \\
    \hline
    \textbf{Spec Constraint} & $\geq 40$dB & $\geq 1$MHz & $\leq 10$mV & $\leq 0.07$V & \\
    \hline
    BO \cite{BO} & 9.66 & \textbf{3.230} & 120.6 & 0.322 & 0.217 \\
    ADO-LLM \cite{ado-llm} & \textbf{55.36} & 0.026 & -3.4 & 0.676 & 0.303 \\
    LLM-USO(R) & 46.78 & 0.067 & 55 & 0.374 & 0.314 \\
    LLM-USO(C) & 44.90 & 0.515 & \textbf{-0.29} & \textbf{0.063} & \textbf{0.383} \\
    \hline
\end{tabular}%
}
\caption{Performance metrics for the hysteresis comparator.}
\label{tab:hyst_comp}
\end{table}

\subsection{Low Dropout Regulator}

We next evaluate the low dropout regulator (LDO) which is used to regulate voltage to a reference of 0.6V. The design space consists of 40 parameters corresponding to the 20 transistors. As shown in Figure \ref{fig:ckt_schematics}a, the LDO circuit has a number of similar sub-structures like the error amplifier, biasing circuit etc. with the two stage differential amplifier and the hysteresis comparator. We show how information reuse from structured design summary of these two circuits can improve the performance metrics of the LDO. As specifications for the LDO, we require the quiescent current to be lower than 12 mA. We define the regulated voltage error as the difference between the reference voltage and the output voltage. We specify for the regulated voltage difference to be less than 0.1V.
There exists a natural trade-off between the current output and the regulation ability. From Table \ref{tab:ldo}, we see that BO does not consider the trade-off between objectives and fails to regulate the output voltage. Due to the larger design space and complex circuit, ADO-LLM prioritizes on voltage regulation at the cost of higher quiescent current. LLM-USO(R) reuses the information from related circuits to understand that both quiscient current and regulated voltage error need to be reduced. LLM-USO(C) explores the LLM suggestions based on uncertainty to output the best design point considering and understanding how to optimize in the presence of tradeoffs.

\begin{table}[!htbp]
\centering
\scriptsize
\resizebox{\columnwidth}{!}{%
\begin{tabular}{|c|c|c|c|}
    \hline
    \makecell{\textbf{Method}} & 
    \makecell{\textbf{Qui. Current}\\\textbf{(mA) $\downarrow$}} & 
    \makecell{\textbf{Regulated Voltage}\\\textbf{Error (V) $\downarrow$}} & 
    \makecell{\textbf{FOM $\uparrow$}} \\
    \hline
    \textbf{Spec Constraint} & $\leq 12$mA & $\leq 0.1$V & \\
    \hline
    BO \cite{BO} & 5.04 & 0.3 & -0.192  \\
    ADO-LLM\cite{ado-llm} & 34.28 & 0.080 & -0.147 \\
    LLM-USO(R) & 10.6 & 0.024 & -0.079 \\
    LLM-USO(C) & \textbf{8.53} & \textbf{0.016} & \textbf{-0.068}  \\
    \hline
\end{tabular}%
}
\caption{Performance metrics for the low dropout regulator.}
\label{tab:ldo}
\end{table}

\subsection{DCDC Converter}

\begin{table}[!htbp]
\centering
\scriptsize
\resizebox{\columnwidth}{!}{%
\begin{tabular}{|c|c|c|c|}
    \hline
    \makecell{\textbf{Method}} & 
    \makecell{\textbf{Output Power}\\\textbf{($\mu W$) $\downarrow$}} & 
    \makecell{\textbf{Converted Voltage}\\\textbf{Error (V) $\downarrow$}} & 
    \makecell{\textbf{FOM $\uparrow$}} \\
    \hline
    \textbf{Spec Constraint} & $\leq 15$ $\mu W$ & $\leq 0.01$V &  \\
    \hline
    BO \cite{BO} & 15.25 & 0.091 & 0.158 \\
    ADO-LLM \cite{ado-llm} & 16.57 & 0.014 & 0.174 \\
    LLM-USO(R) & \textbf{13.13} & 0.006 & 0.185 \\
    LLM-USO(C) & 14.26 & \textbf{0.003} & \textbf{0.199} \\
    \hline
\end{tabular}%
}
\caption{Performance metrics for the DCDC converter. Arrows indicate the desired direction of optimization, and specification constraints are listed for each metric.}
\label{tab:dcdc}
\vspace{-1em}
\end{table}

The DCDC converter is a PWM controlled converter that steps down the input voltage to a lower fixed value. The design space consists of 44 design parameters (consisting of 22 transistors). As seen in Figure \ref{fig:ckt_schematics}b, the DCDC converter has common sub-structures like the the error amplifier, comparator etc with the previously seen circuits. Additionally, it also has some common functionality with the LDO circuit. As a result, we provide design knowledge summaries of the two stage differential amplifier, hysteresis comparator and the LDO circuit for information reuse. As per specifications, we expect the output power to be less than 15 $\mu W$ while the converted voltage error with respect to a reference voltage of 0.6V to be 10mV. From Table \ref{tab:dcdc}, we see that BO fails to meet the design specifications, while ADO-LLM struggles to find the limits till which output power and converted voltage error can be reduced, primarily considering no prior context and the larger design space. LLM-USO(R) is able to more efficiently find smaller output power with reduced converted voltage error. LLM-USO(C) further finds the limits to which both the metrics with inherent tradeoffs can be reduced. 

\section{Conclusion}

This work introduces LLM-USO, a hybrid framework that combines Bayesian Optimization (BO), large language models (LLMs), and structured knowledge representation to tackle the complexities of analog circuit design. By leveraging structured knowledge, LLM-USO facilitates efficient transfer learning, enabling the reuse of design insights from related circuits and significantly enhances optimization efficiency and performance. The key contributions of LLM-USO include a systematic approach for defining, generating, and refining structured knowledge summaries, integrating BO’s uncertainty-based ranking to prioritize high-potential design points, and advancing cross-circuit optimization by reusing insights from similar sub-structures. Experimental evaluations on diverse analog circuits demonstrate that LLM-USO outperforms existing methods, such as ADO-LLM and standard BO. In conclusion, LLM-USO takes a step toward knowledge-driven, efficient, and transferable automated analog circuit optimization akin to human designers.

\section*{Acknowledgment}
This material is based upon work supported by the National Science Foundation under Grant No. 1956313 and by Semiconductor Research Corporation Task No. 3160.055 throughUT Dallas' Texas Analog Center of Excellence (TxACE).

\bibliographystyle{IEEEtran}
\bibliography{main}

\end{document}